\documentclass[aps,prl,superscriptaddress,twocolumn,nofootinbib,floatfix]{revtex4}
\usepackage{amssymb,amsmath}
\usepackage{graphicx,tabularx}

\def\bq{\begin{equation}}
\def\eq{\end{equation}}
\def\ba{\begin{eqnarray}}
\def\ea{\end{eqnarray}}

\newcommand{\etal}{\textit{et al.}}
\newcommand{\ie}{{\sl i.e. }}
\newcommand{\eg}{{\sl e.g. }}

\newcommand{\wt}{\widetilde}

\newcommand{\cp}[1]{\wt{\chi}_{#1}^+}

\newcommand{\cpi}{\wt{\chi}_{i}^+}

\newcommand{\go}{{\wt{g}}}

\newcommand{\ee}{e^+e^-}
\newcommand{\ww}{W^+W^-}

\begin{document}

\title{Breakdown of the Narrow Width Approximation for New Physics}

\author{D. Berdine}
\email{berdine@pas.rochester.edu}
\affiliation{Dept. of Physics and Astronomy, University of Rochester,
             Rochester, NY 14627}

\author{N. Kauer}
\email{kauer@physik.uni-wuerzburg.de}
\affiliation{Institut f\"ur Theoretische Physik, Universit\"at
             W\"urzburg, D-97074 W\"urzburg, Germany}

\author{D. Rainwater}
\email{rain@pas.rochester.edu}
\affiliation{Dept. of Physics and Astronomy, University of Rochester,
             Rochester, NY 14627}

\date{\today}

\begin{abstract}
The narrow width approximation is used in high energy physics to
reduce the complexity of scattering calculations.  It is a fortunate
accident that it works so well for the Standard Model, but in general
it will fail in the context of new physics.  We find numerous examples
of significant corrections when the calculation is performed fully
off-shell including a finite width, notably from effects from the
decay matrix elements, not just phase space.  If not taken into
account, attempts to reconstruct the Lagrangian of a new physics
discovery from data would result in considerable inaccuracies and
likely inconsistencies.
\end{abstract}

\maketitle



The narrow-width approximation (NWA) is used extensively to calculate
cross sections for production of promptly-decaying particles.  It's
use is justified only if five critical conditions are met: (i) the
total width of a resonant particle is much smaller than its mass,
$\Gamma\ll M$; (ii) daughter particles are much less massive than the
parent, $m\ll M$; (iii) the scattering energy is much larger than the
parent mass, $\sqrt{s}\gg M$; (iv) there is no significant
interference with non-resonant processes; and (v) the resonant
propagator is separable from the matrix element (ME).  If these are
valid, the propagator can be integrated independently over all $q^2$
(including unphysical values, with negligible effect) to obtain a
constant:
\bq\label{eq:NWA}
\mathop\int^\infty_{-\infty}
dq^2 \biggl| \frac{1}{q^2-M^2+iM\Gamma} \biggr|^2
\, = \,
\frac{\pi}{M\Gamma}
\eq
In a nutshell, the NWA assumes that the massive state is always
produced exactly at its pole as an asymptotic final state, so its
decay is an independent process, expressed by a simple numerical
constant known as a branching ratio ($BR$): the fractional probability
to decay to a specific final state.  Parametrically, the NWA
introduces an estimated error of ${\cal O}(\Gamma/M)$.  The NWA is
widely used when $\Gamma/M$ is small, regardless of the other
conditions.  While there's some awareness of (i), (iii) and (iv),
assumptions (ii) and (v) have not previously been discussed in the
literature or textbooks to the best of our knowledge.

The NWA became standard at a time when numerical computation tools
were not advanced enough to perform off-shell calculations of full
matrix elements including decays and finite widths.  Complete
analytical calculations for such $2\to n>2$ processes are generally
intractable.  The NWA works well for heavy particle production ($W$
and $Z$ bosons and the top quark, excluding hadronic and flavor
physics) above threshold in the Standard Model (SM), largely owing to
the fact that the decay products are much lighter than the parent.  In
those cases, $\Gamma/M$ is indeed small, a couple of percent at most;
other uncertainties dominate.  We need to include finite widths in
high-energy SM calculations only in a few cases involving threshold
restrictions, such as scattering at the $Z$ pole, $\ee\to\ww\to4f$
production~\cite{RacoonWW}, top quark pair production as a background
at the Large Hadron Collider (LHC)~\cite{OFS-top}, and Higgs boson
decay to weak bosons~\cite{Djouadi:1997yw}.  Our work addresses the
regime above threshold.

In the context of new physics extensions to the SM, it is {\it
generally} the case (but not always) that massive particle widths are
much smaller than their masses, $\Gamma\ll M$, leading one to conclude
that the NWA is still valid.  Also, collision energy is typically far
above production threshold, $E_{\rm CM}-M\equiv\sqrt{s}-M\gg\Gamma$,
thus avoiding a cutoff of the Breit-Wigner lineshape.  We note a
glaring exception to this: the proposed technique to measure new
particle masses via a threshold scan at a future lepton
collider~\cite{Aguilar-Saavedra:2001rg}.  Only one non-NWA study
exists for such a case, and for a small subset of particles in one
model~\cite{FMZ-sleps}.

Rarely does scattering of a given set of initial and final states
result from only one resonant process.  Interference with other
resonant or non-resonant processes can generally occur, rendering the
NWA technically inapplicable.  In the SM at high energy, this is
typically insignificant compared to other uncertainties, but this is
usually not true in new physics scenarios.  We have found numerous
instances of significant corrections from interference effects.
However, these are not the focus of this letter, so we defer that
discussion to a later work~\cite{NWA-full}.

Separation of the resonant particle's propagator is, strictly
speaking, never valid: even when all particles are scalars, as the
phase space factor for the decay particles is a function of $q^2$ for
finite daughter mass, $m\ne0$.  However, for extremely small $m$ the
dependence may be negligible.  The NWA does work very well for top
quark production and decay, despite a non-trivial matrix element and
massive daughter particles.  But in general, momentum-dependent
external wavefunctions and couplings invalidate Eq.~\ref{eq:NWA}.
This can result in significant off-shell corrections in
beyond-the-Standard Model (BSM) scenarios, where the NWA is
universally adopted.  A critical ingredient is massive daughters,
which most BSM scenarios include, often with near-degeneracies driven
by underlying symmetries.  These differences from the SM yield
unanticipated behavior.




\bigskip\bigskip



%
\begin{figure}[ht!]
\includegraphics[width=0.45\textwidth]{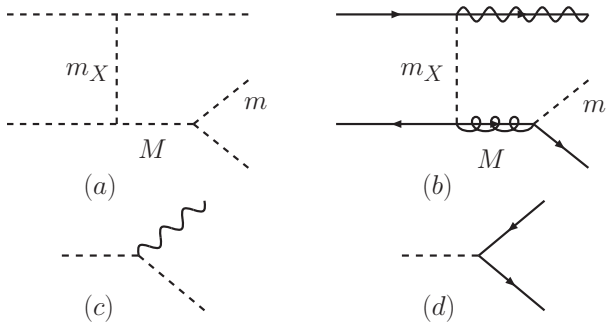}
\caption{Feynman diagrams for various example processes discussed in
the text: (a) scalar field theory toy model; (b) gluino resonance in
supersymmetry; (c) VSS and (d) SFF momentum-dependent renormalizable
interaction vertices.}
\label{fig:Feyn}
\end{figure}

We first consider the process of Fig.~\ref{fig:Feyn}a, $t$-channel
scattering in a pure scalar theory.  It has no matrix elements other
than the $t$- and $s$-channel propagators.  This allows us to study
the $s$-channel particle's off-shell behavior without a decay matrix
element modification.  We may later compare these results with those
of a specific BSM model, where the decay matrix element introduces
additional momentum-dependent structure.  In this toy example we have
only one possible contributing diagram -- other topologies may be
forbidden by assigning flavor or charges to the scalars -- and all
particles are massless except as labeled.  The $t$-channel particle is
kept massive to avoid a forward-scattering singularity, and only one
final-state daughter is massive.  These simplifications allow us to
derive analytical expressions.

Squaring the matrix element and integrating over three-body phase
space obtains the $2\to 3$ off-shell cross section.  Dividing by the
on-shell $2\to 2$ cross section times the na\"ive branching ratio and
subtracting $1$ obtains the deviation of the true leading-order rate
from the NWA result, $\triangle\equiv\sigma_{OFS}/\sigma_{NWA}-1$.  We
defer the rather long full result to later work~\cite{NWA-full}.  Here
we show the result for $\sqrt{s}\gg M$ and finite daughter mass,
$m\ne0$:
\ba\label{eq:SSS-t}
\triangle \, \sim &
- \frac{1}{2}
+ \frac{1}{\pi} \bigg( 1 + \beta^{-2}\frac{\Gamma^2}{M^2} \bigg)
  \tan^{-1} \bigg( \beta^2\,\frac{M}{\Gamma} \bigg)
\\ \nonumber &
- \frac{1}{\pi} \beta^{-2} \frac{\Gamma}{M}
\bigg(
  \frac{M^2}{s}
+ \frac{m^2}{2\,M^2} \,
  \log\left(\frac{M^2(\beta^4\,M^2 + \Gamma^2)}{m^4}\right)
\bigg)
\ea
where $\beta=\sqrt{1-m^2/M^2}$ is the daughter velocity.  The
off-shell result indeed differs from the approximate on-shell result
by a quantity ostensibly of order $\Gamma/M$ (with generally mild
logarithmic corrections), but with an inverse velocity squared that
diverges in the $m\to M$ limit, \ie nearly-degenerate daughter and
parent masses.  This is due to the $q^2$-dependent $t$-channel
propagator and decay phase space factors, and is our first principal
result.  Its interpretation, however, requires some care.

The partial width of a scalar particle decaying to two scalar
daughters is proportional to $\beta^2$, originating from the phase
space.  If this is the only allowed decay, then the total width is
equal to the partial width and the velocity factors cancel.  This
leaves only the log terms as correction factors to a coefficient which
is of order $\alpha_i\times m/M$, where $\alpha_i=g_i^2/4\pi$ is the
coupling strength of the decay interaction.  In such a scenario, the
leading-order off-shell calculation can give ${\cal O}(\alpha_i)$ rate
corrections for decays to a nearly-degenerate daughter, a result which
is interesting in its own right.  The more important implication,
however, is that if multiple decays are allowed, then the velocity
factors for the rarest mode(s) to nearly-degenerate daughter(s) are
not cancelled.

The practical impact is that the effective branching ratio,
$BR_{eff}$, for a rarer mode can be dramatically different than the
na\"ive $BR$, even by an order of magnitude.  If the rare mode is
taggable, as often happens in BSM scenarios, off-shell effects can
alter the phenomenology significantly.  Depending on the relative mass
scales in the problem, the rare mode may be depleted, or enhanced to
some asymptotic finite value in the limit $m\to M$.  This consequence
is our second principal result.  For the $t$-channel process we
considered here, the corrections can even be negative, but this is not
a general rule for arbitrary topologies and particle content.


We don't observe massless scalars in nature, however, so our toy model
is useful only to understand how a calculation performed fully
off-shell can differ from the approximate on-shell result, simply due
to other propagators which depend on $q$.  That the decay threshold
can experience sizeable corrections should not be surprising, as a
production threshold can.  Nevertheless, our result and its impact
does not appear to be known in the literature.  That finite-width
effects can furthermore be enormous for rare decay modes is an
important corollary, easily understood once one obtains the general
analytical form of $\triangle$.  We note that an $s$-channel
all-scalar process results in a slightly different analytical result,
but identical qualitative behavior.

Discoverable BSM physics cannot consist purely of new scalar vertices,
although it may involve additional interactions of a vector and two
scalars (VSS) or a scalar and two fermions (SFF), shown in
Figs.~\ref{fig:Feyn}c and~\ref{fig:Feyn}d.  Both types introduce
additional momentum dependence into the matrix element.  The decay
matrix element for scalar decay to fermions (S:FF) is proportional to
$q^2-(m_1+m_2)^2$, where $q^2$ is the invariant mass of the resonance
and $m_i$ are the final-state fermion masses.  The VSS vertex is
proportional to the difference of the scalar particles' momenta, in
turn roughly proportional to $q$.  In general, the decay matrix
element alters the integration over $q^2$, rendering the NWA formally
invalid.  Our task is to see how much the off-shell result can differ
from the na\"ive one.



We move on from a simple toy model to a realistic new-physics scenario
and examine its practical phenomenology.  For this purpose we choose
the minimal supersymmetric Standard Model (MSSM) as our framework.
Supersymmetry (SUSY) is a highly-motivated scenario, and there are
several readily-available tools for performing off-shell calculations
in the MSSM~\cite{Reuter:2005us}.  The MSSM spectrum contains one new
particle of opposite spin statistics for every particle degree of
freedom in the Standard Model, in addition to two Higgs doublets.  If
supersymmetry exists, however, it is a broken symmetry, as any SUSY
particles that might exist must be much more massive than their SM
partners.  We might expect to find SUSY particles at several hundred
GeV in mass, but not much above the TeV scale.  Many SUSY scenarios,
such as anomaly-mediated SUSY breaking (AMSB), have natural
near-degeracies in their spectrum, driven by high-scale vacuum
expectation values for fields, which generate the low-scale physical
spectrum.  Nearly-degenerate sparticles would still appear in cascade
decays of heavier sparticles to the lightest sparticle, the dark
matter candidate.  This landscape is ripe for off-shell effects.

One MSSM process which has no possible non-resonant interference
diagrams is $u\bar{d}\to\cpi\go$ ($i=1,2$), with gluino decay to
strange quark plus squark, as shown in Fig.~\ref{fig:Feyn}b.  As a
mixed EW--QCD process, at LHC it would be challenging to dig out from
QCD SUSY processes, unless SUSY were realized in a long-lived chargino
scenario such as AMSB.  It is however an excellent proxy for
demonstrating the physics inherent in off-shell resonance effects.
That is, $\wt{q}\go$ production would exhibit a very similar matrix
element effect, but the presence of QCD non-resonant interference
would muddy the present lesson.

Analytically at leading order in $\Gamma/M$ and $1/s$ where allowed,
the off-shell to NWA cross section deviation is:
\begin{multline}\label{eq:FFS_GP}
\triangle \, \sim \,
-\frac{1}{2} + \frac{1}{\pi}
\Bigg(1 + \frac{(M^2-2m^2)\,\Gamma^2}{\beta^4\,M^4}\Bigg)
\tan^{-1} \Bigg( \beta^2\frac{M}{\Gamma} \Bigg)
\\
+ \frac{1}{2\pi} \frac{\Gamma}{M} \beta^{-4}
\cdot 
\Bigg(
  \frac{m^4}{M^4}
  \ln\Bigg(\frac{\beta^4\,M^4 + M^2\Gamma^2}{m^4}\Bigg)
\\
- \ln\Bigg(\frac{\beta^4\,M^4 + M^2\Gamma^2}{s^4/m_X^4}\Bigg)
\Bigg)
\end{multline}
\noindent
with parent/daughter notation as before.  In the limits $m\to0$ and
$\Gamma\ll M$, the first two terms go to 0, and the remaining
correction is ${\cal O}(\Gamma/{M})$, as expected.  There is also a
residual $\log(s)$ dependence, which would not be noticed unless far
above production threshold.

We immediately see an important difference compared to the all-scalar
toy model, which has no decay matrix element: the $\Gamma/M$
corrections are proportional to $\beta^{-4}$ instead of $\beta^{-2}$
(differences in log terms are of minor importance).  This comes from
the additional powers of $q$ from the decay matrix element.  This is
our third principal result.  We thus expect even more dramatic
corrections to effective branching ratios for rare decays to
nearly-degenerate daughters in spectra with dominant decays to lighter
particles.  These lighter particles dominate the total width, so
$\Gamma_{\rm tot}$ is independent of $\beta^4$.  It should be clear
that the radically different behavior of off-shell BSM resonances is
also due to final-state masses not significantly smaller than the
resonance mass.

We define $BR_{eff}$ as the off-shell cross section to the final state
of interest, divided by the sum of all possible final state cross
sections, each also calculated off-shell.  For our example, this is
three-body production of the final state $\cp1\bar{s}\wt{s}_{L,R}$.
We choose the MSSM parameter space benchmark point SPS1a~\cite{SPS},
which also allows for gluino decays to lighter top quark plus stop,
and bottom quark plus sbottom; these together generate a gluino
partial width of 2.6~GeV, $\sim0.4\%$ of the 600~GeV gluino mass.  A
typical width-to-mass ratio for MSSM gluinos lighter than about 1~TeV
is $1-5\%$.

\begin{figure}[ht!]
\includegraphics[width=0.42\textwidth]{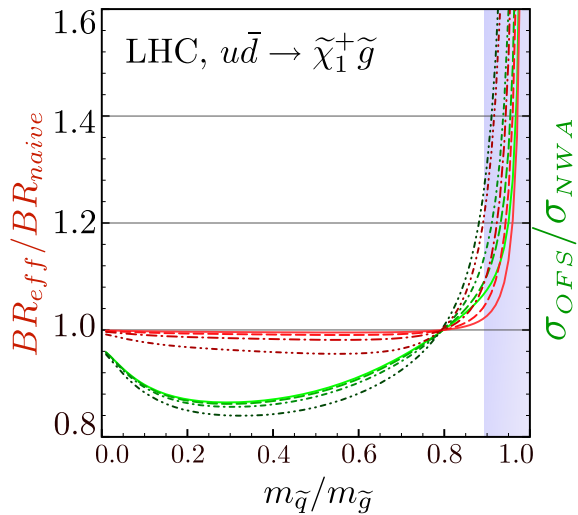}
\caption{Ratio of effective to na\"ive BRs (left axis) and off-shell
to NWA cross sections (right axis) for the MSSM process
$u\bar{d}\to\cp1\bar{s}\wt{s}_L$ at the LHC
(cf. Fig.~\protect\ref{fig:Feyn}(b)).  The resonsant gluino has an
additional partial width of $0.5,1,2,5\%$ of its mass (solid, dashed,
dashdotted and dotted curves) due to decays to stops and sbottoms.
The first- and second-generation squarks lie in the shaded band for
all SPS benchmark points with a heavier gluino.}
\label{fig:BReff}
\end{figure}

Numerical results for $BR_{eff}$ are shown in Fig.~\ref{fig:BReff} for
a scan over daughter squark mass (we assume degenerate 1st- and
2nd-generation squarks).  The shaded band delineates the region
containing all 1st- and 2nd-generation squarks in all SPS benchmark
scenarios with a heavier gluino.  As expected, if the squarks are
lighter than the stops and sbottoms, they dominate the total width, so
$BR_{eff}\approx1$.  However, note that for squarks much lighter than
the gluino, the {\it cross section} receives sizeable corrections.
This tree-level effect on the overall normalization is at a level
comparable to the QCD next-to-leading order uncertainty.

For $\wt{s}_L$ within about $10\%$ of the gluino mass, $BR_{eff}$ can
greatly exceed the na\"ive NWA expectation.  Where the $BR$ becomes
small, if it is a taggable rare mode (which most decays would be),
then these corrections become important, lest we extract incorrect
Lagrangian parameters from the relative branching ratios observed in
data.  At SPS1a, this decay mode would receive almost a $25\%$
correction, an order of magnitude larger than the $\Gamma/M$ estimate,
and greater than the residual QCD next-to-leading order production
rate uncertainty.  The NWA-derived ratio of this mode's branching
ratio relative to (unaffected) decays to lighter sbottoms would
disagree with the mass spectra measured via other
methods~\cite{edges}.  Results are nearly identical for decays to
$\wt{s}_R$.

\begin{figure}[ht!]
\includegraphics[width=0.4\textwidth]{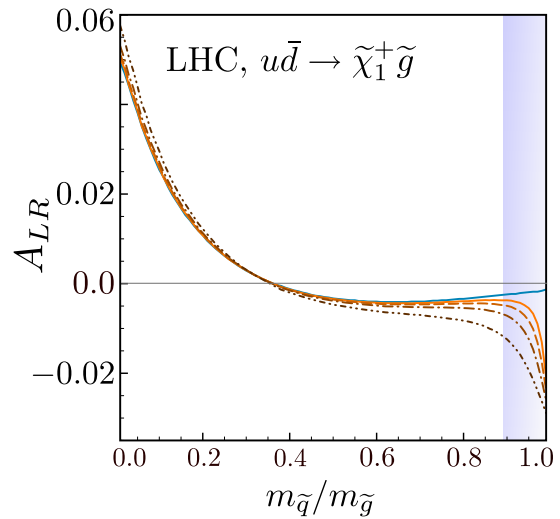}
\caption{MSSM gluino decay asymmetry to degenerate-mass left-chiral 
v. right-chiral squarks as a function of the squark to gluino mass
ratio, in $\cp1\go$ production at the LHC, as discussed in the text.
The NWA predicts exactly zero for all masses.  The yellow (blue)
curves are for multi-mode (single-mode) decays; line types and shaded
band as in Fig.~\protect\ref{fig:BReff}.}
\label{fig:asym}
\end{figure}

In addition to effective branching ratio corrections, we observe
chirality selection in gluino decays.  In the NWA, the chiral-blind
Majorana-fermion gluino will decay with equal rates to left- and
right-chiral squarks and anti-squarks, given equal masses (that is, up
to phase space effects for the small electroweak mass splittings which
typically appear).  However, the initial- and final-state helicities
are connected via the Dirac structure of the fermion chain.  The
production side of the event in our example involves a chargino, which
selects left helicity.  The gluino's fermion mass can flip the
helicity, but does not give equal rates.  This is our fourth principal
result.  We show it graphically in Fig.~\ref{fig:asym}.

While squark masses below a couple hundred GeV are already ruled out
by Tevatron searches, we see that there would be a few-percent
asymmetry for nearly-degenerate squark masses.  Left-chiral squarks
prefer to decay to charginos, while $SU(2)_L$-singlet right-chiral
squarks cannot.  These final states differ qualitatively and can be
distinguished, but a detailed study is necessary to determine what
level of asymmetry would be observable.

We have examined multiple other cases involving VSS and SFF vertices.
Examples include sbottoms decaying to stop plus $W$ boson (S:SV), and
squarks decaying to quark plus gluino (S:FF).  Their general
$\sigma_{OFS}/\sigma_{NWA}$ behavior is qualitatively similar to what
we find for the F:FS case, again depending on $\beta^{-4}$, with minor
variations in the log terms and coefficients.  Most MSSM particle
decays would exhibit the same phenomenological features of our primary
example.

%



In summary, we investigated off-shell matrix element effects in
scattering processes involving new heavy states which decay to other
massive states.  In general, if multiple decay modes are allowed, the
accuracy of the effective branching ratios differs from the na\"ive
${\cal O}(\Gamma/M)$ expectations based on the NWA, even by orders of
magnitude.  Rarer modes may receive enormous corrections.  If
neglected, this would corrupt the extraction of model parameters from
data.  Additionally, massive Majorana fermions exhibit a helicity
selection effect which may introduce observable asymmetries into the
data.  Unanticipated, such asymmetries would likely be incorrectly
interpreted as signals of additional new physics, such as CP
violation.  Our results are based on a neglected but important aspect
of tree-level cross section calculations.

Our results are far more generally applicable than just to
supersymmetry.  They are based on general matrix element behavior for
arbitrary renormalizable interactions, phase space and integration of
a heavy resonance's propagator over a range in $q^2$ for decay to
massive daughter particles.  For instance, Universal Extra Dimensions
models~\cite{Macesanu:2005jx} are another example of cascade decays
and very close degeneracies.


\end{document}